\documentclass[aps,prl,twocolumn,groupedaddress,10pt]{revtex4-1}

\usepackage{amsmath,amssymb,bm,graphicx,graphics}

\newcommand{\tmin}{{\theta\text{-}\hspace{-1pt}\min}}

\newcommand{\supsection}[1]{\vspace{5mm}\begin{center}\textbf{\uppercase{#1}}\end{center}}

\begin{document}

\title{Predicting the speed of epidemics spreading on networks}
\author{Sam Moore}
\email{s.moore@bath.ac.uk}

\author{Tim Rogers}
\email{t.c.rogers@bath.ac.uk}
\affiliation{Centre for Networks and Collective Behaviour, Department of Mathematical Sciences, University of Bath, Bath, BA2 7AY, UK}

\begin{abstract}
Global transport and communication networks enable information, ideas and infectious diseases now to  spread at speeds far beyond what has historically been possible. To effectively monitor, design, or intervene in such epidemic-like processes, there is a need to predict the speed of a particular contagion in a particular network, and to distinguish between nodes that are more likely to become infected sooner or later during an outbreak. Here, we study these quantities using a message-passing approach to derive simple and effective predictions which are validated against epidemic simulations on a variety of real-world networks with good agreement. In addition to individualized predictions for different nodes, we find an overall sudden transition from low density to almost full network saturation as the contagion develops in time. Our theory is developed and explained in the setting of simple contagions on tree-like networks, but we are also able to show how the method extends remarkably well to complex contagions and highly clustered networks. 
\end{abstract}

\maketitle

It took more than nine years for the Black Death to spread across Europe. Progress of this devastating outbreak of bubonic plague was limited by 14$^{\textrm{th}}$ century travel networks to an average daily dispersion of approximately 1.5km \cite{mcevedy1988bubonic}. In frightening contrast, the recent Zika outbreak in South America was found to spread with an average daily dispersion of 42km, rising as high as 634km in the most densely populated parts of Brazil \cite{zinszer2017zika}. This extraordinary difference is indicative of a mobile society that is no longer rigidly bound by spatial structure, making the relevant notion of distance network-based rather than geographic. Similarly, in the highly connected  domain of social media, the spread of concepts, memes and hashtags can be explosive. One recent empirical study of the dynamics of online rumour cascades --- often reaching tens of thousands of users in a matter of days --- made the worrying finding that false information spreads faster then true \cite{vosoughi2018spread}. It takes little imagination to see how an understanding of propagation speeds in modern networks would have, in the digital case, great commercial and political benefit, and in the physical case be invaluable in planning outbreak prevention, monitoring and response. 

The field of network epidemiology \cite{moore2000epidemics,keeling2005networks,danon2011networks,Pastor2015} has developed a wide spectrum of techniques for the analysis of spreading processes. One approach to the problem of spreading speed is through numerical simulations (see e.g. \cite{van2011gleamviz}), which yield useful results on small scales, but for increasingly large complex networks may prove slow and impractical. Alternative approximations have been made by considering only the most probable path between a given target node and the source \cite{gautreau2008global}. It is known that this shortest-path approach can significantly overestimate the infection arrival times  \cite{gautreau2007arrival}, but to  take into account all possible paths would soon be infeasible as their number typically grows exponentially with the number of  vertices in the network. One promising idea is a conjectured connection between centrality measures and infection arrival time \cite{borgatti2005centrality}, which so far has only been tested numerically.

While global networks of interest are highly connected, they are also typically sparse in the sense that individuals usually interact with a number of others that is very small relative to the total population size. Exploitation of this sparse network structure has been a key tool in network epidemiology, in particular via the message-passing approach pioneered in \cite{karrer2010message}. This technique has allowed for efficient characterization of the epidemic (percolation) threshold \cite{Hamilton2014,Karrer2014}, and gave rise to the new notion of \emph{non-backtracking centrality} \cite{martin2014localization}. In \cite{rogers2015assessing,kuhn2017} a message-passing approach was used to make individualized predictions for node responses to spreading processes, giving a physical interpretation of non-backtracking centrality as the probability for a node to appear in the percolating cluster. None of these works has yet addressed the important questions of how fast an epidemic will spread in a given network, and which nodes may fall victim first. 

Here, we seek to assess the full time-dependence of an epidemic outbreak in order to characterize the speed of spread in a given network by calculating the mean delay in infection between nodes at different graph distances from the source. Technically, we achieve this through a saddle-point analysis of the left tail of the distribution of time to infection, expressed via the message-passing equations. This method enables us to find the overall speed of an infection in a network, and to show that the arrival time at a node is accurately predicted by the logarithm of its non-backtracking centrality. 

Our theoretical predictions for both spreading speed and arrival times show excellent agreement with numerical simulations performed on real-world networks, even in the case of highly clustered contact networks with heavy tailed degree distributions. Remarkably, we show that the method can also be extended to complex threshold models of contagions in which a node must be exposed to multiple infective neighbours before acquiring the contagion itself. We finish by observing that the time for the infection to spread through the bulk of the network is independent of network size, implying an almost instantaneous jump from low to high density of infection when time is properly scaled; a propoerty which we show to be common to time-ordered percolation in general.

\textbf{\textit{Speed of spread.}} We begin by considering a simple SI infection spreading on a sparse network starting from a single infected node (details on the extension to other models, inlcuding SIR and complex contagions, are found in the supplement). When node $i$ becomes infectious, it transmits the infection to a neighbour $j$ after a delay $X_{i\to j}$; a random variable drawn from a distribution with density $f(x)$, independently from any other event. The choice of an exponential distribution for $f$ would correspond to Markov disease dynamics, although it has been shown that real-world contagion dynamics differ substantially from this simple case \cite{gough1977estimation,anderson1980spread,lloyd2001realistic,eichner2003transmission}, and hence we study general distributions of transmission time. 

Write $T_i^n$ for the length of the shortest (temporal) path to a node at distance $n$ from $i$, and $T_{i\to j}^n$ for the shortest such path whose first step is to node $j$. It follows that $T^n_i=\min_{j\in\partial i}\,T_{i\to j}^n$, where $\partial i$ denotes the set of neighbours of $i$. More generally, $T_{i\to j}^n$ decomposes as
\begin{equation}
T_{i\to j}^n=X_{i\to j}+\min_{k\in\partial j\setminus i}T_{j\to k}^{n-1}\,.
\label{Ti}
\end{equation}
Writing $F_{i\to j}^n(t)$ for the probability that $T_{i\to j}^n$ is less than $t$, we arrive at the message passing equation
\begin{equation}
F_{i\to j}^n(t)=\int_0^t f(x) \bigg(1-\prod_{k\in\partial j\setminus i}\left[1-F_{j\to k}^{n-1}(t-x)\right]\bigg)\,\textrm{d}x\,.
\label{mpF}
\end{equation}
In writing the above we have assumed independence between the variables $\{T_{j\to k}^{n-1}\}$; although this technically only holds for tree graphs, we will see that the approximation is effective for a broad class of real-world networks. 

Equation (\ref{mpF}) represents a nested hierarchy of expressions which could in principle be solved numerically for a given network, infection and source node. However, this process is computationally intensive and the results are not generalisable. We will pursue a different path and investigate the structure of the dynamics described by (\ref{mpF}) to reveal useful general insights.

\begin{figure}
\includegraphics[width=\linewidth, trim=40 0 50 0, clip]{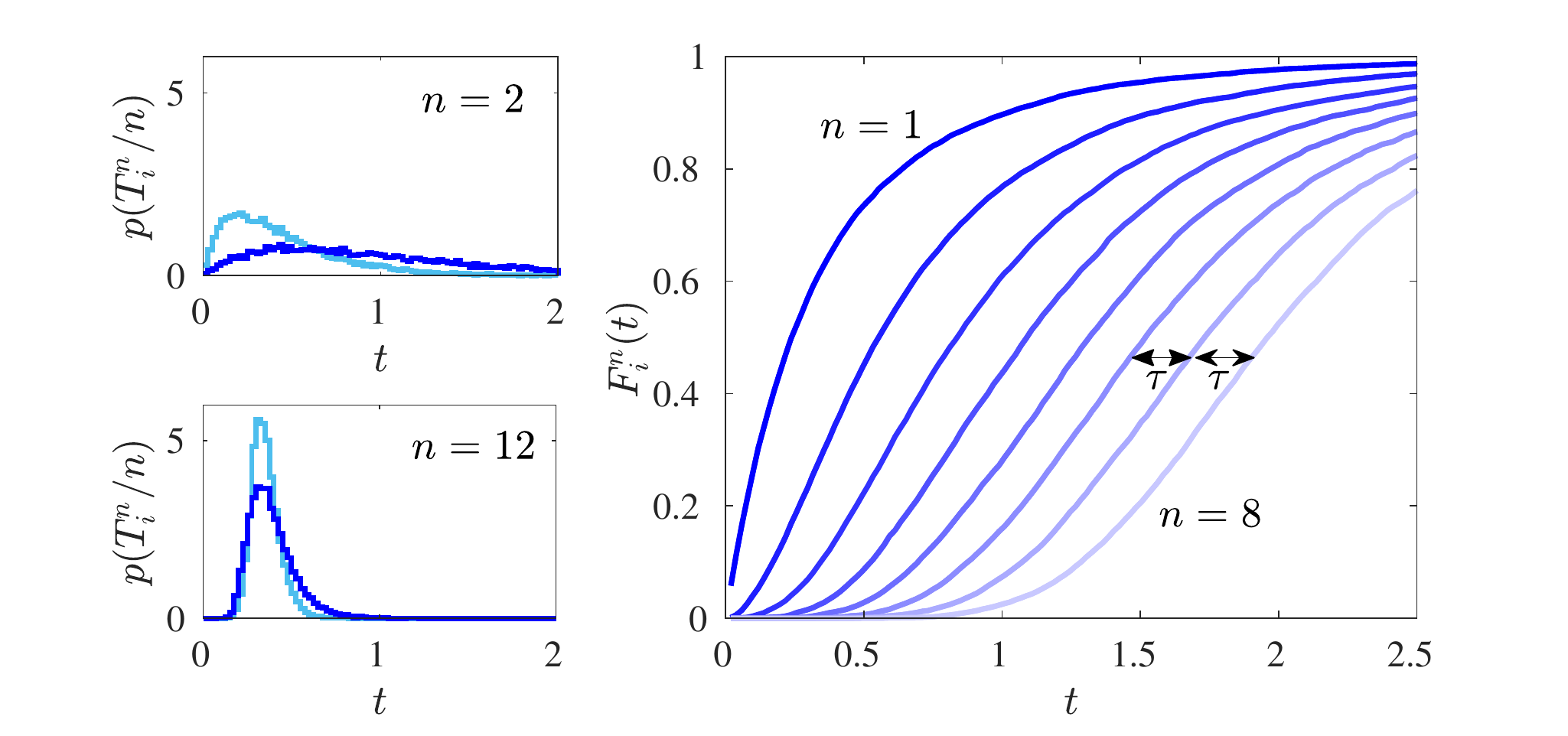}
\caption{Left panels: simulation of the distribution of the scaled time $T_i^n/n$ for an epidemic to reach distance $n$ from a source node $i$ chosen to have degree 1 (dark) or degree 3 (pale); as $n\to\infty$ these distributions will converge to delta functions at some value $\tau$. Right: simulation of the CDF $F_i^n(t)$ for time to reach distance $n$ from a source node $i$ chosen to have degree 1, showing convergence to a standard form with a fixed offset $\tau$. In both cases node-to-node transmission times are standard exponentials and the network is an Erd\H{o}s-R\'enyi graph with mean degree 3 on $N=10^4$ nodes.}
\label{tau_n_regular}
\end{figure}

At first glance, it appears that the spreading process depends in a complicated way on the precise layout of the network, however, we find that the system possesses a regularity which emerges after a few iterations. In a network of $N\gg 1$ nodes, for $1\ll n \ll N$ we observe the convergence $T_{i}^n/n \to \tau$ for some constant $\tau$, describing the delay between spreading $n-1$ steps from the source to $n$. In this sense $1/\tau$ can be interpreted as the \textit{speed} of spreading in the network. This effect is illustrated in the left panels of Fig.~\ref{tau_n_regular}, showing the convergence and reduction of variance in simulated histograms of $T_{i}^n/n$ for different source nodes $i$ as $n$ grows.

To compute the characteristic delay $\tau$, we examine the left tails of $F_{i\to j}^n$ for large $n$. Our rationale for this approach is that, as illustrated in Fig.~\ref{tau_n_regular}, the offset is the same across the whole distribution, and we will show that the left tails are amenable to a linear analysis. For $t\ll n\tau$, we linearize (\ref{mpF}) to obtain
\begin{equation}
F_{i\to j}^n(t)\approx \int_0^t f(x) \sum_{k\in\partial j\setminus i}F_{j\to k}^{n-1}(t-x)\,\textrm{d}x\,.
\label{LmpF}
\end{equation}

This problem is mathematically analogous to that of front propagation and we therefore follow the standard method described in \cite{van2003front}. The trivial solution $F^0(t)\equiv0$ is linearly unstable with increasing $n$, and the dominant rate of growth will determine $\tau$. The two sided Laplace transform of Eq.~(\ref{LmpF}) reads
\begin{equation}
\tilde F_{i\to j}^n(k)=\tilde f(k) \sum_{k\in\partial j\setminus i} \tilde F_{j\to k}^{n-1}(k)\,,
\label{lin}
\end{equation}
where $ \tilde f(k)=\int e^{-kx}f(x)\textrm{d}x$ is the Laplace transform of $f$. Viewing $\tilde F$ as a vector with entries indexed by directed edges, Eq.~(\ref{lin}) describes an iterative process of multiplying by a matrix that encodes the entries of the sum, and then by the scalar $\tilde{f}(k)$. Thus, for large $n$ we can expect   
\begin{equation}
\tilde F_{i\to j}^n(k)\propto v_{i\to j}\, e^{-\omega (k) n},
\label{anz}
\end{equation}
where the coefficient $v_{i\to j}$ contains the edge-specific information, and the function $ \omega (k)$ determines the overall exponential growth rate. Substituting this ansatz into (\ref{lin}), we find $\bm{v} =\tilde f(k)e^{\omega (k)}B\bm{v}$, where $B$ is the non-backtracking matrix \cite{martin2014localization}. This is an eigenvalue equation for $B$ with a non-negative eigenvector $\bm{v}$; according to the Perron-Frobenius theorem, for a connected network there is a unique maximum eigenvalue $\lambda$, which is real and positive. Thus the growth rate is found as $\omega(k)= -\log(\lambda \tilde f(k))\,.$
Note that $1/\lambda=\rho_c$ is the percolation threshold of the network \cite{Hamilton2014,rogers2015assessing,kuhn2017}.

Examining the inverse transform at time $t+n\tau$, one finds (full details are in the supplement) physically meaningful results in the limit of large $n$ only when 
\begin{equation}\label{taumaster}
\tau=\max_k\bigg\{\frac{1}{k}\big(\log\rho_c  - \log\tilde{f}(k)\big) \bigg\}\,.
\end{equation}
This is our first main result, showing how the speed of spread is determined by the network via its percolation threshold $\rho_c$, and by the infection itself via the Laplace transform of its transmission time distribution. It is important to note that this result is derived from making a tree-like assumption for the underlying network, and our calculation holds in the limit of large distance from the source. In this sense it describes the fastest spreading regime; the mid-outbreak phase of exponential growth. 

\begin{figure}
\includegraphics[width=0.95\linewidth]{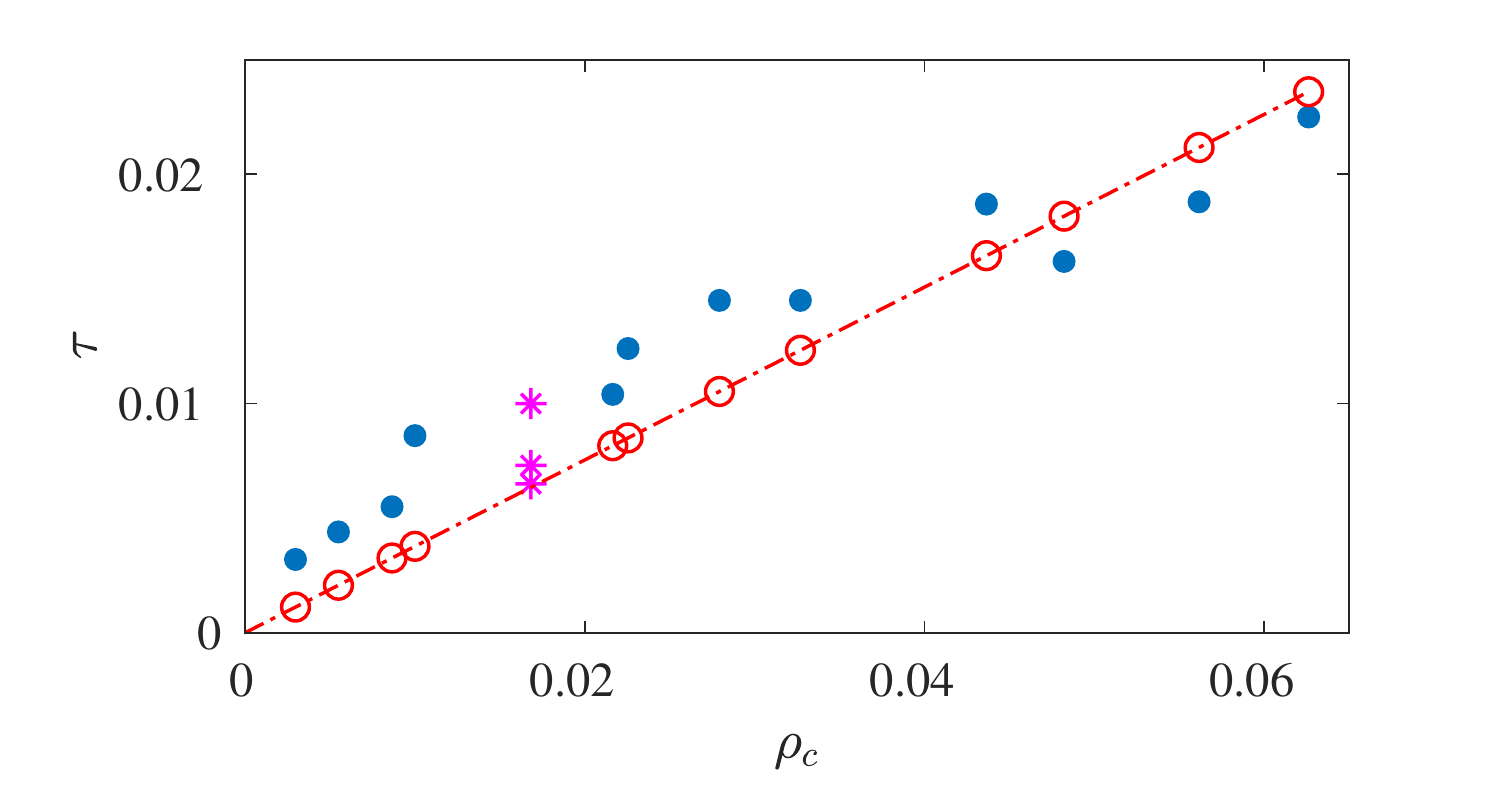}
\caption{Predicted and observed values of the spreading delay $\tau$ for a unit rate exponential infection spreading on a variety of social and communication networks \cite{rozemberczki2018gemsec,leskovec2007dynamics,yang2015defining,leskovec2007graph,cho2011friendship,leskovec2009community,richardson2003trust,salathe2010high} with different percolation thresholds $\rho_c$. Predicted values (red circles) are calculated using Eq.~\eqref{taumaster}. Observed values (blue dots) show the average over $10^3$ simulations with random source nodes. Stars show results for Watts-Strogatz random graphs on $10^4$ nodes with degree 30 and rewiring probabilities of 0.1, 0.5 and 1. Full details of all simulations are given in the suplement.}
\label{fig:taunets}
\end{figure}

In practical applications, however, most networks of interest are not tree-like, and finite size effects mean the infection is unlikely to be able fully accelerate to the stable regime we have calculated. Nonetheless, our result still provides high-quality predictions. Fig.~\ref{fig:taunets} demonstrates the effectiveness of this measure on a variety of real world networks from the Stanford Large Network Dataset Collection (SNAP) \cite{snapnets}, many with heavy-tailed degree distributions and high clustering; Table~S.I in the suplement gives full details. To further test the reliance of our method on the tree-like assumption made in writing (\ref{mpF}), we have simulated spreading processes in Watts-Strogatz random graphs with varying rewiring probabilities. Included in Fig.~\ref{fig:taunets}, the results for these networks show that our method performs better for higher rewiring probability, but is still very successful for highly clustered networks with low rewiring. 

\begin{figure}
\includegraphics[width=0.95\linewidth]{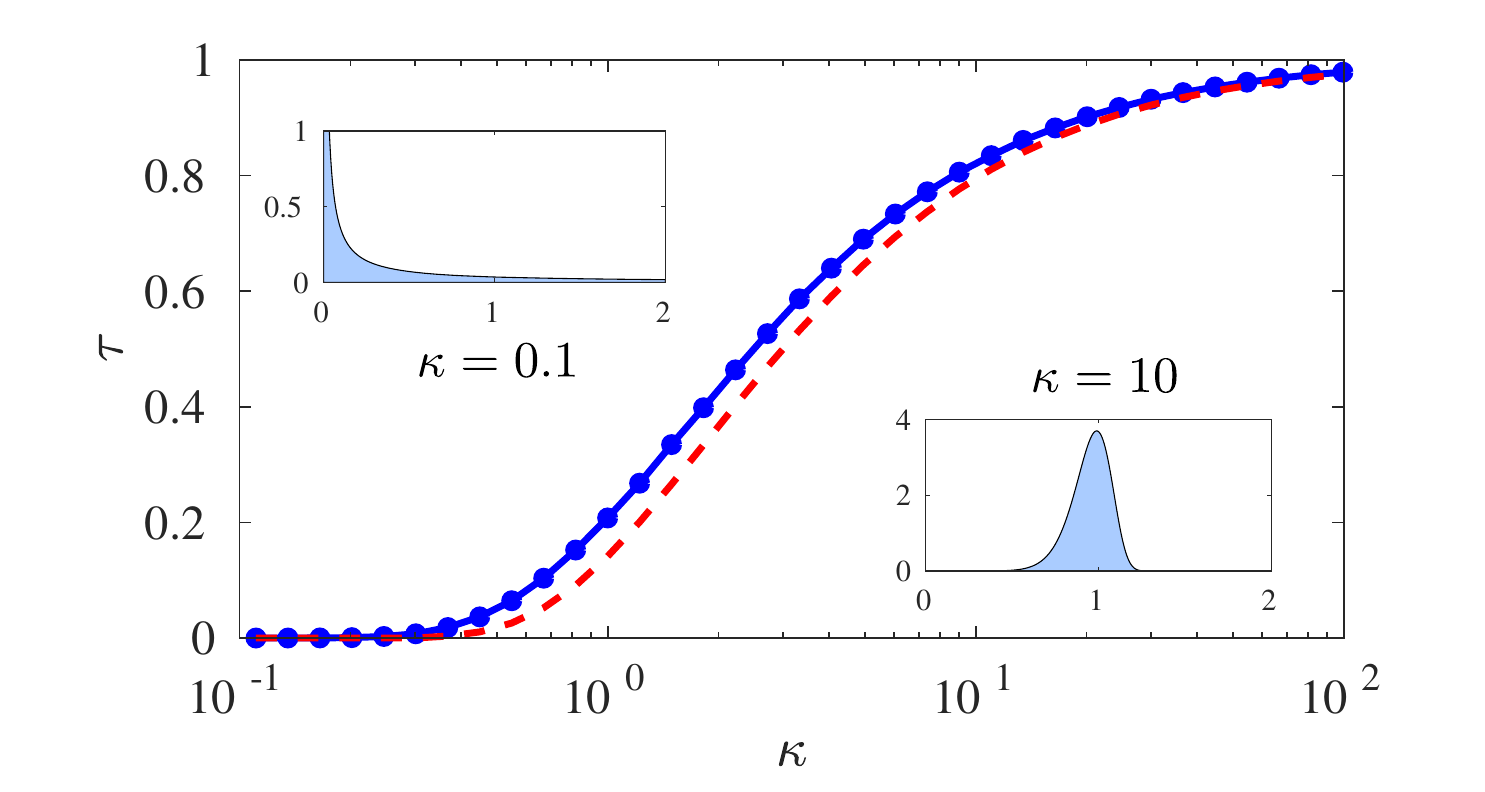}
\caption{Simulated (blue marked line) and predicted (red dashed line) spreading delay $\tau$ for infections with Weibull delay times with varying shape parameter $\kappa $ and fixed mean 1 (example delay distributions shown in insets). Simulations follow the same method as Fig.~\ref{fig:taunets}, averaged over 100 samples on an Erd\H{o}s-R\'enyi graph of $10^4$ nodes and mean degree 3.}
\label{fig:tau-v-k}
\end{figure}

As well as the network, our measure of speed also depends on properties of the infection. 
One might expect the time delay $\tau$ to be scaled by the mean delay time, but beyond this it is difficult to discern from (\ref{taumaster}) how the shape of the distribution should affect the global speed of spread. To explore this aspect we show in Fig.~\ref{fig:tau-v-k} the observed and predicted spreading speed for Weibull distributed delays, interpolating between heavy-tailed and Dirac distributed. Crucially, we find that the shape of the distribution of transmission time has a substantial effect on the speed of spread in a network. If there is mass near zero then delays are minimal due to the presence of extremely fast transmission routes. Conversely, if transmission time is close to deterministic then spreading is determined entirely by graph distance, meaning $\tau\approx1$. In the supplement we prove that $\tau$ is always less than the mean delay time, with equality only for Dirac-delta distributions. 

\begin{figure}[t!]
\includegraphics[width=0.24\textwidth]{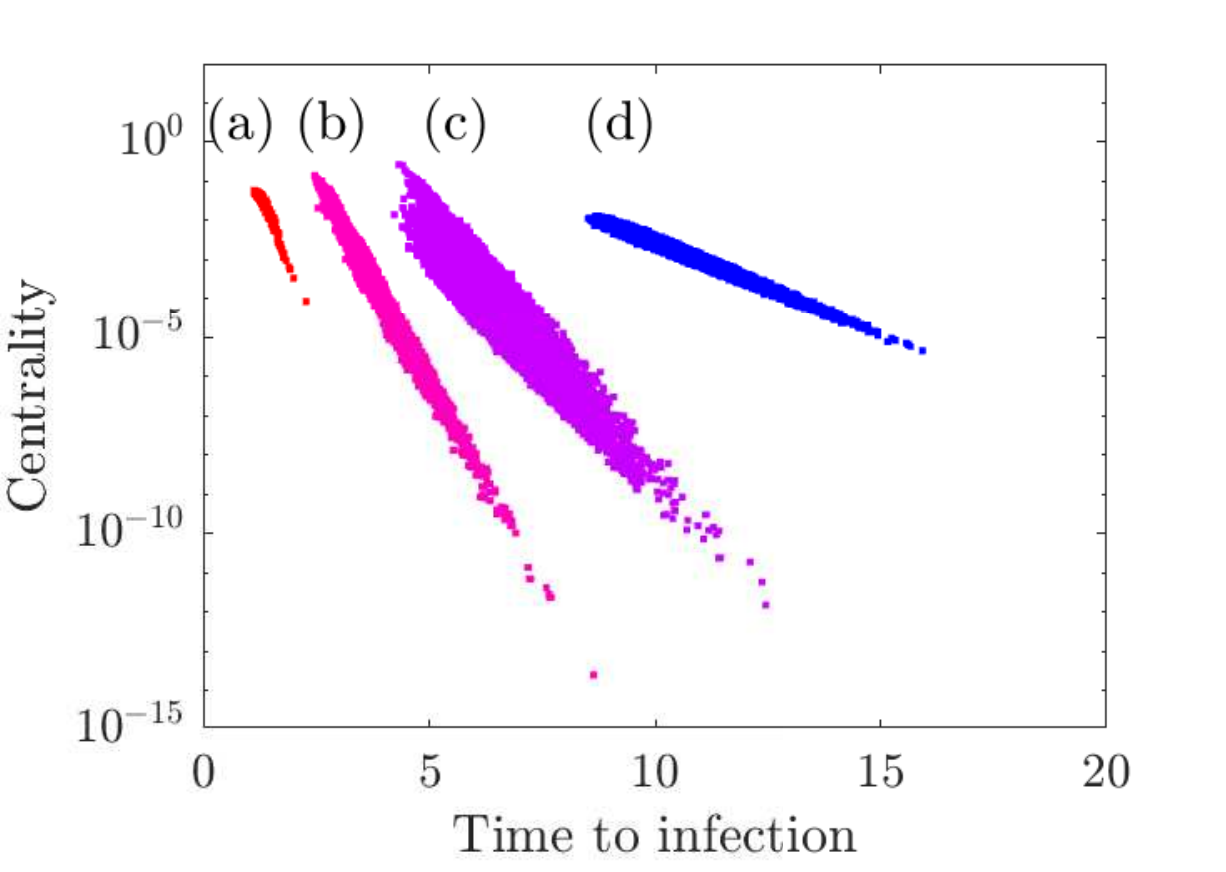}\includegraphics[width=0.24\textwidth]{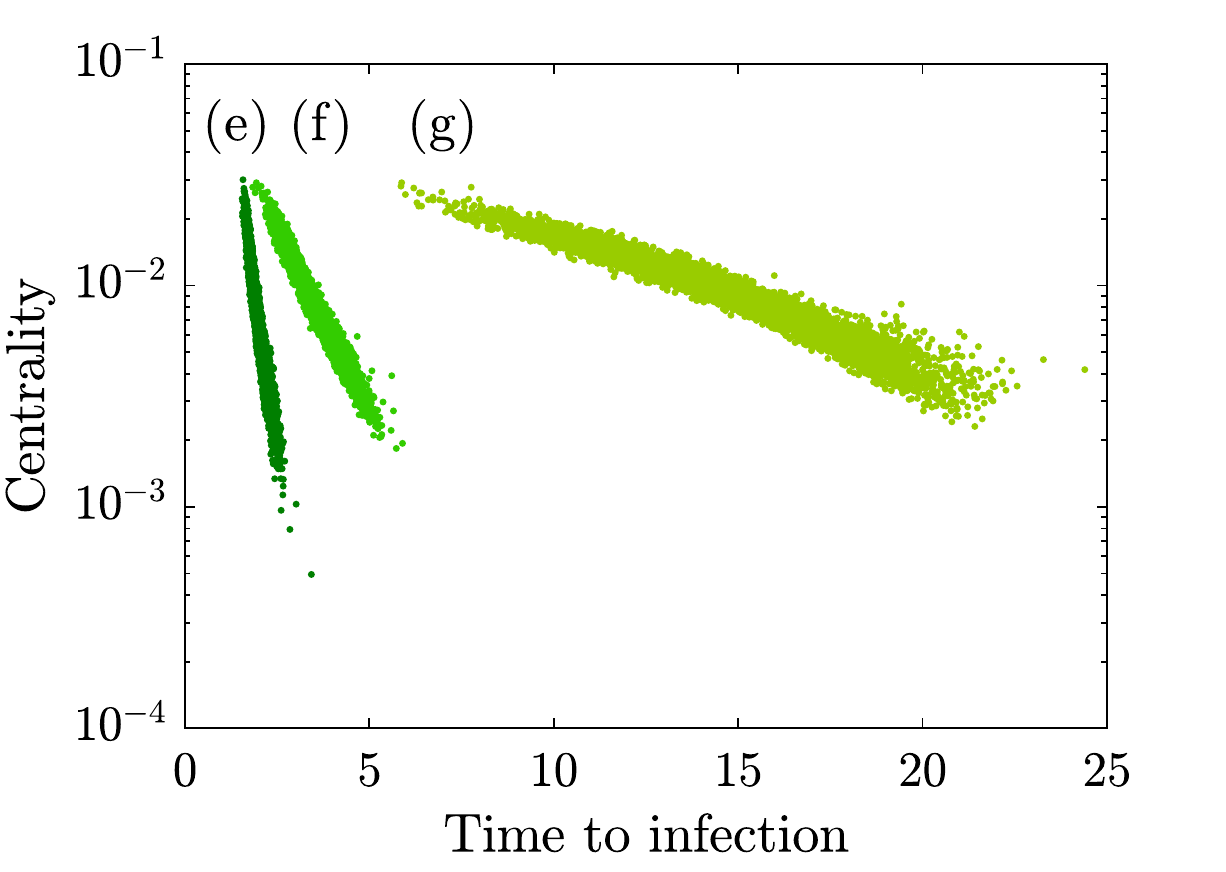}
\caption{Non-backtracking centrality predicts time to infection. Left: scatter plot of centrality and average arrival time for nodes in a selection of networks using a contagion with Weibull ($\kappa=10$) infection times. Right: results for simulations of complex contagions with threshold $\theta$ in an Erd\H{o}s-R\'enyi graph with $10^4$ nodes and mean degree 6. Key: (a) an inter-personal contact network in an American high school \cite{salathe2010high}, (b) `Epinions' social media, (c) `Deezer' Romanian social network \cite{rozemberczki2018gemsec}, (d) an Erd\H{o}s-R\'enyi graph with $N=10^5$ nodes, (e) $\theta=1$, (f) $\theta=2$, (g) $\theta=3$.}
\label{infvtime}
\end{figure}

\textbf{\textit{The time taken to receive the infection.}} As well as predicting the overall spreading speed, our approach also allows us to rank nodes in the network by their expected time to become infected. Write $\Delta_{ij}$ for the offset in infection time between nodes $i$ and $j$, which for large $n$ should satisfy $F^n_i(n\tau)=F^n_j(n\tau+\Delta_{ij})$. Inverting the transform in Eq.~(\ref{anz}) for large $n$ by steepest descent and comparing with the above (details in the supplement) we find that 
\begin{align}
\Delta_{ij}=\frac{1}{k^\star}\log\left(\frac{c_i}{c_j}\right)+\mathcal{O}(1/n)\,,
\label{Delta}
\end{align}
where $k^\star=\textrm{argmax}_k \{\omega(k)/k\}$ and $c_i=\sum_{j\in\partial i}v_{i\to j}$ is the non-backtracking centrality of node $i$. This log-linear relationship is demonstrated numerically in Fig.~\ref{infvtime} for nodes in a selection of networks from SNAP. This result is important as it resolves the open question of exactly how network centrality measures may be used to estimate epidemic arrival time, and provides a robust theoretical justification for the use of non-backtracking centrality (see supplement for a comparison to other centrality measures).

\begin{table}
\resizebox{\columnwidth}{!}{
\begin{tabular}{|l|c|c|c|c|c|c|}
\hline 
 & \multicolumn{3}{c|}{\textbf{Exponential}} &\multicolumn{2}{c|}{\textbf{Weibull} ($\kappa=10$)} \\\hline
\textbf{Network name} & \textbf{Simple} &\textbf{2-core} & \textbf{3-core} & \textbf{Simple}& \textbf{2-core} \\\hline
Erd\H{o}s-R\'enyi &-0.9487&-0.9693&-0.9502& -0.9721&-0.9691\\
Epinions \cite{richardson2003trust}&-0.8765&-0.7769&-0.7428&-0.9946&-0.8661\\
Deezer Croatia \cite{rozemberczki2018gemsec}&-0.8265&-0.8094&-0.8088&-0.9049&-0.8697\\
Facebook Artists \cite{rozemberczki2018gemsec}&-0.7943&-0.7481&-0.7322&-0.9586&-0.8778\\
Arxiv Cond. Mat.\cite{leskovec2007graph}&-0.8712&-0.7999&-0.7939&-0.9513&-0.8296\\
Facebook Companies \cite{rozemberczki2018gemsec}&-0.8439&-0.7203&-0.6685&-0.9138&-0.7606\\
School contact \cite{salathe2010high}&-0.7667&-0.7532&-0.7199&-0.9661&-0.9371
\\\hline
\end{tabular}
}
\caption{Correlation coefficient between contagion arrival time (measured from $10^3$ simulated spreading processes with random sources) and the logarithm of non-backtracking centrality, for various networks. Values close to the theoretical limit $-1$ correlation imply strong prediction quality. }
\label{tab}
\end{table}

Going further, many realistic models of network contagion require the number of infected neighbours of a node to reach some threshold $\theta\geq1$ before the infection is passed on. In the supplement, we show how a variation of our theory, building on results from \cite{shrestha2014message}, extends to these complex contagion models by considering the $\theta$-shortest temporal paths from a node. Remarkably, the log-linear relationship derived above continues to hold in this more complex setting, as illustrated in Fig.~\ref{infvtime}. In addition to this visual demonstration, we present in Table~\ref{tab} the Pearson correlation coefficients between log-non-backtracking centrality and infection arrival time for various disease dynamics in various networks. These results show that our theory, which is physically justified and cheap to compute, provides excellent predictions of the relative delay between nodes in a wide variety of spreading processes.

\begin{figure}[t!]
\includegraphics[width=0.85\linewidth, trim=50 0 60 0]{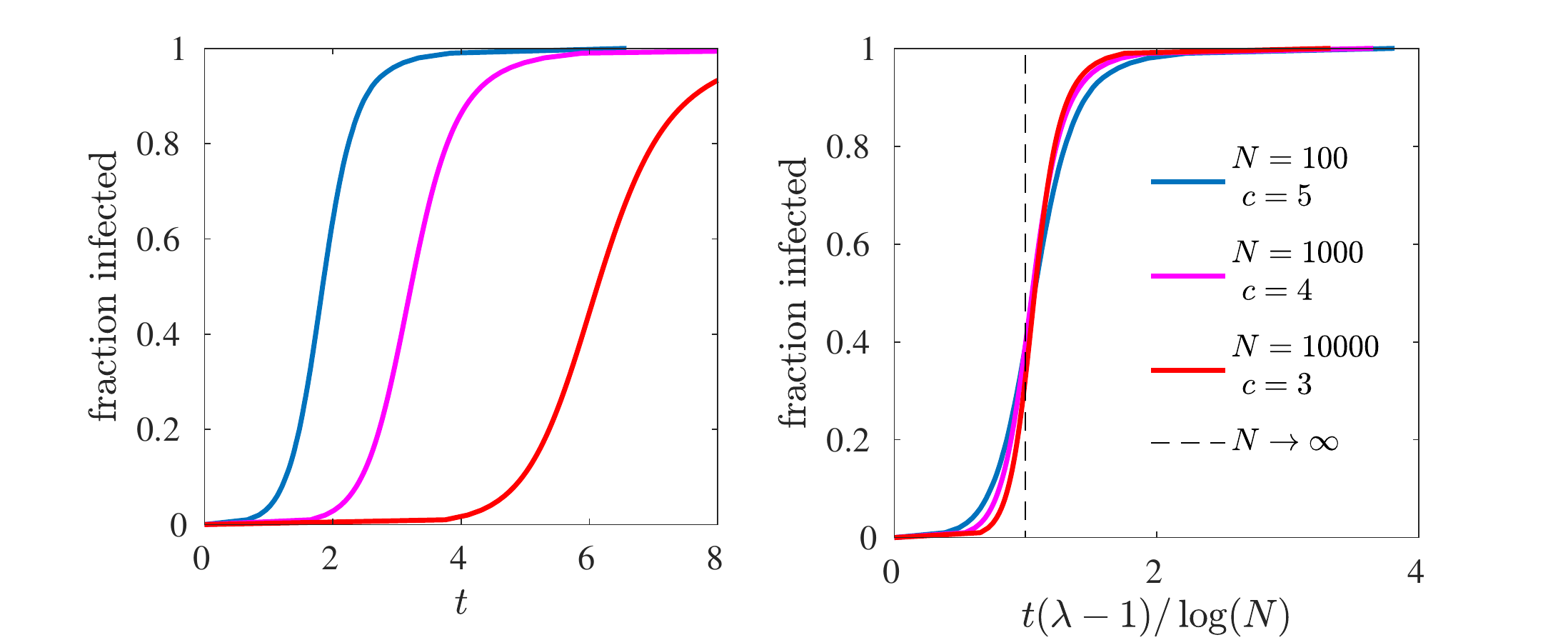}
\caption{Fractional size of the cluster of infected node as a function of time in various Erd\H{o}s-R\'enyi graphs of different sizes $N$ and mean degree $c$, averaged over 100 simulations from random seed nodes, with standard exponential infection times. The left panel shows real time, the right has rescaled time showing convergence to a step function in the limit $N\to\infty$ implying `instantaneous' spread to the bulk of the network.}
\label{fig:trans-sparse}
\end{figure}
Since the non-backtracking centrality of a node is mainly a property of its local environment, the result (\ref{Delta}) means that we should expect the vast majority infections to occur during a time window whose duration is independent of the total size of the network. However, it can be shown that in a network of size $N$ the time needed for an infection to take hold grows like $\log(N)/(\lambda-1)$. Taken together these results imply that, on the timescale of the spreading contagion in a large network, one will observe an almost instantaneous jump between a vanishing fraction of nodes infected to almost complete infection. We illustrate this result in Fig.~\ref{fig:trans-sparse} for Erd\H{o}s-R\'enyi graphs of increasing size, and provide precise theoretical derivations in the supplement, where we show that this property holds for models of temporal percolation in both sparse and dense networks. 

\textbf{\textit{Discussion.}}
We have presented here a theoretical framework for determining the speed of contagion processes in large networks. Analysing the spreading front of contagion probability we derived Eq~(\ref{taumaster}), showing how network topology and infection dynamics affect speed via, respectively, the network percolation threshold and the Laplace transform of the transmission time law. Our theory also reveals in Eq.~(\ref{Delta}) a surprisingly simple relationship between contagion arrival times and the non-backtracking centrality of nodes. Finally, we have observed that these results imply that spreading process in large networks undergo an almost instantaneous expansion in their reach when time is properly scaled. 

The setting for our theoretical derivation has been that of simple epidemics spreading on large tree-like networks. However, we have shown that the key results hold remarkably well for a broad class of networks, including those with high clustering, and for contagion models including non-Markov dynamics and complex threshold models. Further development of rigorous mathematical results for these models is a challenging problem worthy of considerable future efforts. Excitingly, our results suggest possible routes for the development of monitoring and intervention protocols for real-world contagions using message-passing methods. Progress in this direction may require the consideration of even more detailed models including temporally varying and multi-layered networks; both promising avenues for future research. 

\textit{Acknowledgements.} TR was supported by The Royal Society, SM was supported by a scholarship from the EPSRC Centre for Doctoral Training in Statistical Applied Mathematics at Bath (SAMBa), under the project EP/L015684/1.

\bibliography{spread}

\newpage
\onecolumngrid

\begin{center}
\huge Supplemental material: \\Predicting the speed of epidemics spreading on networks
\end{center}
\large

\supsection{Calculation of time delay}

\begin{figure}[h]
\includegraphics[width=0.5\linewidth]{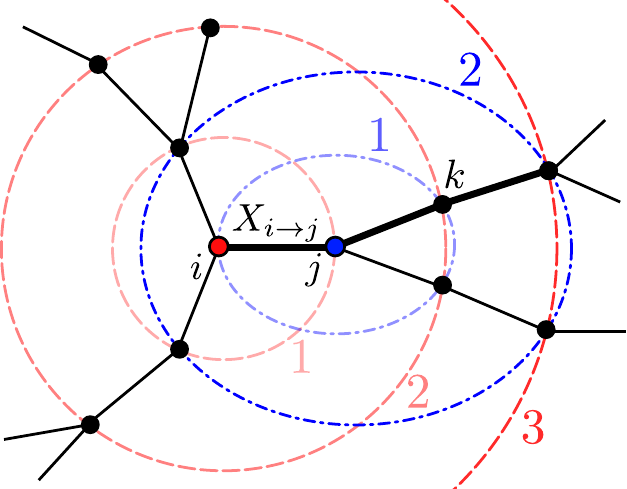}
\caption{The shortest temporal path (bold) from node $i$ to reach distance $n$ ($=3$ here) going via $j$ has length given by the delay $X_{i\to j}$, plus the minimum length of a path reaching distance $n-1$ from $j$ that does not go via $i$.}
\label{drawing}
\end{figure}

Recall that we write $T_i^n$ for the shortest (temporal) path to a node at distance $n$ from $i$, and $T_{i\to j}^n$ for the shortest such path whose first step is to node $j$. As illustrated in Fig.~\ref{drawing}, we can decompse these quantities as follows:
\begin{equation}
T^n_i=\min_{j\in\partial i}\,T_{i\to j}^n\,,\quad T_{i\to j}^n=X_{i\to j}+\min_{k\in\partial j\setminus i}T_{j\to k}^{n-1}\,,
\label{Ti2}
\end{equation}
where $\partial i$ denotes the set of neighbours of $i$. 

Writing $F_{i\to j}^n(t)$ for the cumulative distribution function of $T_{i\to j}^n$, Eq.~(\ref{Ti2}) implies the message passing equation
\begin{equation}
F_{i\to j}^n(t)=\int_0^t f(x)\left[1- \prod_{k\in\partial j\setminus i}\left(1-F_{j\to k}^{n-1}(t-x)\right)\right]\,\textrm{d}x\,.
\label{mpF2}
\end{equation}
For $t\ll n\tau$, we linearize (\ref{mpF2}) to obtain
\begin{equation}
F_{i\to j}^n(t)\approx \int_0^t f(x) \sum_{k\in\partial j\setminus i}F_{j\to k}^{n-1}(t-x)\,\textrm{d}x\,.
\label{LmpF2}
\end{equation}
The two sided Laplace transform of Eq.~(\ref{LmpF2}) reads
\begin{equation}
\tilde F_{i\to j}^n(k)=\tilde f(k) \sum_{k\in\partial j\setminus i} \tilde F_{j\to k}^{n-1}(k)\,,
\end{equation}
where $ \tilde f(k)=\int e^{-kx}f(x)\textrm{d}x$ is the Laplace transform of $f$. For large $n$ we can expect   
\begin{equation}
\tilde F_{i\to j}^n(k)\propto v_{i\to j}\, e^{-\omega (k) n}.
\label{anz2}
\end{equation}
Here $v$ is identified as the top eigenvector of the non-backtracking matrix $B$ with entries
\begin{equation}
B_{i\to j,k\to\ell}=\begin{cases}1&\text{if}\, k=j\,\text{and}\,\ell\neq i\\0&\text{else.}\end{cases}
\end{equation}
Using the ansatz (\ref{anz2}), the inverse transform reads 
\begin{equation}
F_{i\to j}^n(t)=\frac{v_{i\to j}}{2\pi \text{i}}\int_{c-i\infty}^{c+i\infty}\varphi(k)e^{kt-\omega(k)n}\,\textrm{d}k\,,
\end{equation}
where $ c $ is chosen freely in the region of convergence of $ \tilde F_{i\to j}^n $.

To have convergence of time delay to a constant $ \tau $, setting $ t=t+n\tau $ to get
\begin{equation}\label{ntau}
F_{i\to j}^n(t+n\tau)=\frac{v_{i\to j}}{2\pi \text{i}}\int_{c-i\infty}^{c+i\infty}\varphi(k)e^{kt-(\omega(k)-\tau k)n}\,\textrm{d}k\, ,
\end{equation}
we should expect not to see either exponential growth or decay in $ F $. Now for large $n$ the integral \eqref{ntau} will be dominated by the saddlepoint $ k^{\star}=c+iz^{\star} $. So then the requirement for  $ F $ exhibiting neither exponential growth nor decay means that the modulus of the exponential factor in \eqref{ntau} must vanish at this point.

That is for $ k^{\star} $ such that
\begin{align}\label{kstar}
\left. \frac{\text{d} ( \omega(k)-\tau k)}{\text{d}k}\right|_{k^{\star}}=0
\end{align}
then we require that
\begin{align}\label{tau}
\Re[\omega(k^{\star})]-\tau c =0 \,  .
\end{align}
Combining these we get
\begin{align}\label{mintau}
\tau=\omega'(k^{\star})=\frac{\Re[\omega(k^{\star})]}{c} \, ,
\end{align}
implying $ \Im[\omega'(k^{\star})]=0 $. 
Further since
\begin{align}
&\omega'(k^{\star})=\frac{\Re[\omega(k^{\star})]}{c} \iff \left. \frac{\text{d} ( \omega(k)/k)}{\text{d}k}\right|_{k^{\star}} =0
\end{align}
then the possible $ k^{\star} $ are the stationary points of $ \frac{\omega(k)}{k}  $ with $ \tau $ the value at these points. Hence choosing $ k^{\star} $ to maximize $ \tau $ finds the asymptotic  time delay we are seeking.

\supsection{An upper bound on delay time}
We claim that $ \tau \leq \mathbb{E}X $ where $ X\sim f $, the random variable for transmission time. The proof of this follows from considering our expression for  $ \tau= \min_{k}\left[ \frac{-\log(\lambda \tilde{f}(k))}{k}\right]  $. The Laplace transform of $ f $ is 

\begin{align}
\tilde{f}(k)=\mathbb{E}(e^{-kX})  \text{ where } X \sim f \, ,
\end{align}
then by Jensen's inequality, since $ e^{-x} $ is convex,
\begin{align}
\mathbb{E}(e^{-kX}) \geq e^{-k\mathbb{E}(X)}
\end{align}
with equality if $f$ is the Dirac delta distribution.
Then
\begin{align}
-\frac{1}{k}\log (\lambda\tilde{f}(k)) &\leq -\frac{1}{k}\log (\lambda e^{-k\mathbb{E}(X)}) =  \mathbb{E}(X) -\frac{1}{k}\log (\lambda)
\end{align}
and 
\begin{align}
\tau \leq \max_{k}\left[  \mathbb{E}(X) -\frac{1}{k}\log (\lambda)\right] = \mathbb{E}(X) 
\end{align}
which is attained for  Dirac delta  $f$. Therefore, as noted in the main text, we expect time delay to be minimal for heavy tailed $ f $ and maximal, going to $ \mathbb{E}(X)  $, for Dirac delta $ f $. 

\supsection{Infection time offset}
To quantify the heterogeneity in infection recieval we take advantage of the symmetry of the infection process to write
\begin{align}\label{offset}
F^n_i(n\tau)=F^n_j(n\tau+\Delta_{ij}) \, .
\end{align}
Thus an approximation of the offset, $ \Delta $, may be found by first approximating the c.d.f 
\begin{align}
 F^n_i(t) = 1-\prod_{j\in\partial i}\left( 1-F^n_{i\to j}(t)\right)  \, .
\end{align}

Now using \eqref{kstar} and \eqref{tau} it follows that

\begin{align}
\omega(k)-\tau(k) \approx (\Im[\omega(k^{\star})]-\tau z^{\star})i -\frac{1}{2}\omega''(k^{\star})(k-k^{\star})^2
\end{align}

and thus taking $ \Delta k=k-k^{\star} $ equation \eqref{ntau} gives us that

\begin{equation}
F_{i\to j}^n(t+n\tau)\approx \frac{v_{i\to j}}{2\pi i}\int_{c-i\infty}^{c+i\infty}\varphi(k)e^{\phi}\,\textrm{d}k
\end{equation}
where 
\begin{align}
\phi=(k^{\star}+\Delta k)t-(\Im[\omega(k^{\star})]-\tau z^{\star})in-\frac{1}{2}\omega''(k^{\star})\Delta k^2n  \,.
\end{align}
Further taking $ \frac{1}{2}\omega''(k^{\star}) =D$, then
\begin{align}
F_{i\to j}^n(t+n\tau) \approx&  \frac{v_{i\to j}}{2\pi i}  e^{-(\Im[\omega(k^{\star})]-\tau z^{\star})in+k^{\star}t}\int_{c-i\infty}^{c+i\infty}\varphi(k)e^{-Dn(\Delta k-\frac{t}{2Dn})^2+\frac{t^2}{4Dn}}\,\textrm{d} k \,.
\end{align}
As $ n $ becomes  large then the integrand becomes dominated by the contribution at $ k^{\star} $ and we may approximate to a Gaussian integral to find
\begin{align}
F_{i\to j}^n(t+n\tau)\mathrel{\mathop{\approx}\limits_{n\to\infty}}   -\frac{v_{i\to j}\varphi(k^{\star})}{2\sqrt{\pi Dn}}  e^{-(\Im[\omega(k^{\star})]-\tau z^{\star})in+k^{\star}t+\frac{t^2}{4Dn}}
\end{align}
Going back to equation (\ref{Ti2}) we then have, for $t\ll n\tau$,
\begin{align}
F^n_i(n\tau)=& 1-\prod_{j\in\partial i}\big(1-F^n_{i\to j}(n\tau)\big) \approx  \sum_{j\in\partial i}\big(F^n_{i\to j}(n\tau)\big) \\\nonumber
\approx& -\frac{\varphi(k^{\star})}{2\sqrt{\pi Dn}}  e^{-(\Im[\omega(k^{\star})]-\tau z^{\star})in}\sum_{j\in\partial i}v_{i\to j}\,.
\end{align}
Thus, substituting this approximation into \eqref{offset} and solving gives
\begin{align}
\Delta_{ij}=\frac{1}{k^\star}\log\left(\frac{c_i}{c_j}\right)+\mathcal{O}(1/n)\,,
\end{align}
where $k^\star=\textrm{argmax}_k \omega(k)/k$ and $c_i=\sum_{j\in\partial i}v_{i\to j}$ is the non-backtracking centrality of node $i$

\supsection{Different disease dynamics}
It is straightforward to adapt our set-up to include SIR models by adding a time $R_i$ to recovery, so that transmission from $i$ to $j$ only occurs if $X_{i\to j}<R_i$. This presents the interesting complication that there is generally a difference between the time to transmit an infection up to a certain distance from a source, versus the time to recieve an infection that starts a certain distance away. This effect was explored in detail in \cite{rogers2015assessing}. However, for the asymptotic study of the fastest infection routes (obtained by considering the linearisation of the message passing equations), this distinction is not important. Similarly, SIS and SIRS dynamics do not behave substantially differently in this regard, since only the fastest passage time is relevant, not the subsequent recovery and reinfection dynamics. 
\begin{figure}
\centering
\includegraphics[width=0.7\linewidth, trim=80 10 80 0]{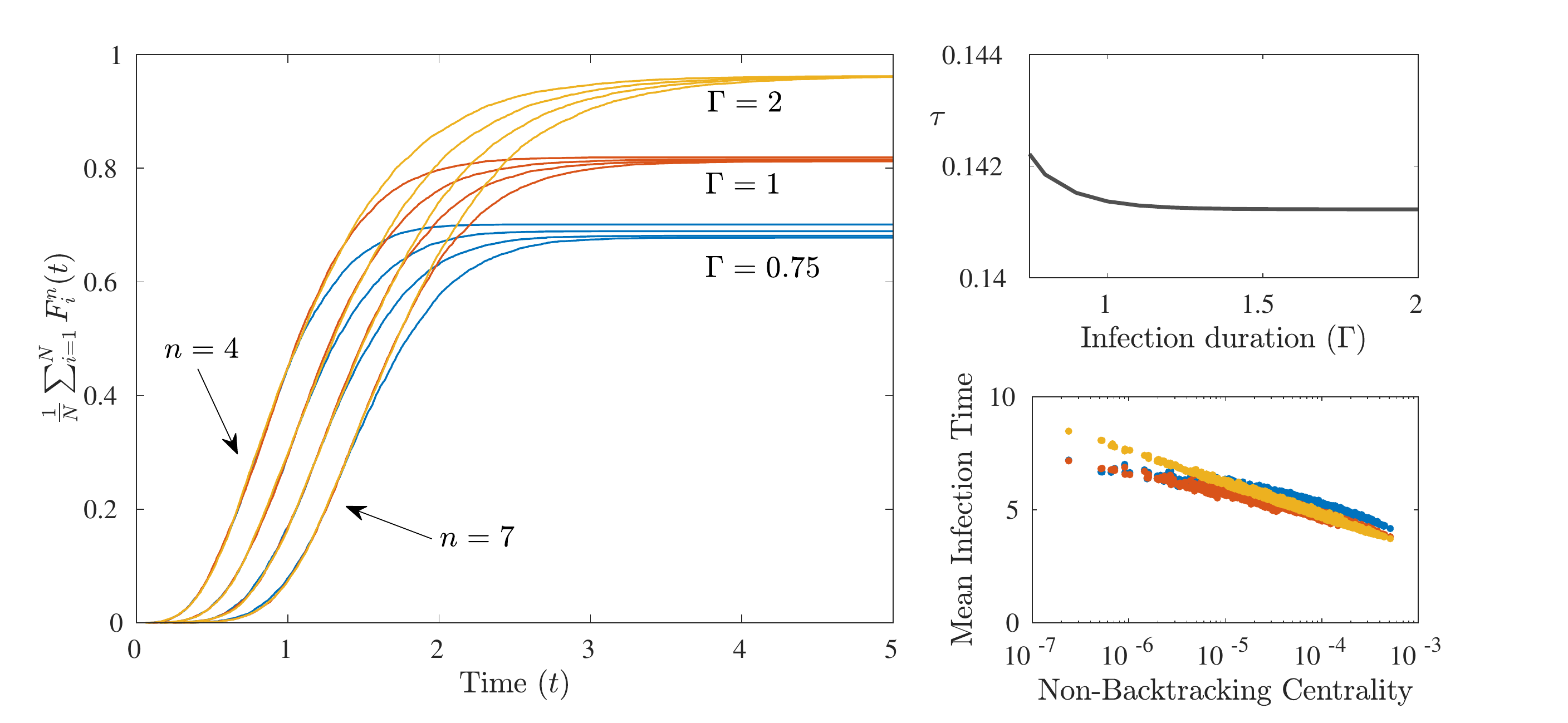}
\caption{Spreading speed of the SIR model with fixed infectious period $\Gamma$ and unit rate (i.e. $\beta=1$) infection. Left: simulation results on the cumulative distribution function of time to reach a given distance $n$, averaged over all starting points, for different durations of infection. The uncertain transmission affects the final fraction of contagions reaching a given distance, but does not appreciably affect the speed. Right upper: analytical calculation of the delay $\tau$ as a function of $\Gamma$, showing very little variation. Right lower: scatter plots of the correlation between node non-backtracking centrality and mean time to infection (excluding contagions that do not reach the node). We used an Erd\H{o}s-R\'enyi random graph of $N=10^4$ and mean degree $c=3$.}
\label{fig:sir}
\end{figure}

In general, uncertain transmission is incorporated to our settting simply by allowing $X_{i\to j}=\infty$ with some probability, so that $\int_0^\infty f(x)\text{d}x=\rho<1$. Here $\rho$ corresponds to the edge occupation probability in the bond percolation model induced by the epidemic \cite{moore2000epidemics}. In Figure~\ref{fig:sir} we show example results for the SIR model in which individuals remain infected for a fixed duration $\Gamma$, during which they infect their neighbours at constant rate $\beta$. For this example, the transmission time density function is $f(x)=\beta e^{-\beta x}\mathbb{I}_{x<\Gamma}$. The corresponding edge occupation probability is $\rho=1-e^{-\beta\Gamma}$ and Laplace transformed density is $\tilde{f}(k)=\frac{\beta}{\beta+k}(1-e^{-(k+\beta)\Gamma})$. As expected, varying the infectious period $\Gamma$ has almost no effect on the speed of propagation of the contagion, which is determined by the fastest transmission route and hence unaffected if slower routes are trimmed. 

More interestingly, many popular models of contagion on networks require the number of infected neighbours of a node to reach some threshold $\theta\geq1$ before the infection is passed on. Our formalism extends to these complex contagion models by considering the $\theta$-shortest temporal paths from a node to the bulk of the network.

Specifically, the time that node $i$ becomes infected in a complex contagion on a large network with threshold $\theta$ maps to the large $n$ limit of $T_{i}^n$ defined by 
\begin{equation}
T^n_i=\tmin\{\,T_{i\to j}^n\,|\,j\in\partial i\}\,,
\end{equation}
where ``$\tmin$'' denotes the $\theta$-smallest element of the specified set, and 
\begin{equation}
T_{i\to j}^n=X_{i\to j}+\tmin\{T_{j\to k}^{n-1}\,|\,k\in\partial j\setminus i\}\,.
\end{equation}

As detailed in \cite{shrestha2014message}, the corresponding equations for the cumulative density functions are
\begin{equation}
F_{i\to j}(t)=\int_0^t f(x) \left[1-\sum_{\stackrel{\scriptstyle M\subseteq \partial j\setminus i}{|M|<\theta}}\,\,\prod_{k\notin M}\left(1-F_{j\to k}(t-x)\right)\prod_{m\in M}F_{j\to m}(t-x)\right]\,\text{d}x\,.
\label{tF}
\end{equation}
Note that the special case $\theta=1$ corresponds to the simple contagion model previously considered, and indeed the inner sum in Eq.~(\ref{tF}) contains only the element $M=\emptyset$, and hence reduces to Eq.~(2).

Our strategy for analysis in the $\theta=1$ case was to consider the left tails of $F$, in which the recursion equation can be linearised. The simple physical intuition for this linear theory is that for a node to receive the infection unusually early, it is only necessary (and indeed likely) for one neighbour to be infected. Unfortunately, for the case of $\theta\geq2$, such a linear theory is not possible as early infection of a node requires $\theta$ of its neigbours to be infected early. Mathematically, this rule is manifested in the fact that the small $F$ expansion of the right hand side of (\ref{tF}) has order $\theta$. 

An alternative approach is to consider the recursion map $\mathcal{G}$ defined by the action
\begin{equation}
\mathcal{G}[\bm{F}^n](t)=\bm{F}^{n+1}(t-\tau)\,,
\end{equation}
that is, $\mathcal{G}$ maps the collection of functions $\{F^n_{i\to j}\}$ to their updated versions according to equation (\ref{tF}), offest by the (as yet unknown) spreading delay $\tau$. In the limit of many applications of $\mathcal{G}$ we have convergence 
\begin{equation}
\mathcal{G}^n[\bm{F}]\stackrel{n\to\infty}{\longrightarrow}\bm{F}^\star\,,
\end{equation}
where $\bm{F}^\star$ is a non-trival limiting profile function (i.e. not identically one or zero). Let us consider the asymptotic stability of $\mathcal{G}$ around $\bm{F}^\star$. Let $e$ be a directed edge in the network, then
\begin{equation}
\begin{split}
\frac{\partial \mathcal{G}[\bm{F}]_e(t)}{\partial F_{e'}(t')}
&=-\int_0^{t-\tau} f(x) \sum_{\stackrel{\scriptstyle E\subseteq \partial e}{|E|<\theta}}\frac{\partial}{\partial F_{e'}(t')}\left[\prod_{e''\notin E}\left(1-F_{e''}(t-\tau-x)\right)\prod_{e''\in E}F_{e''}(t-\tau-x)\right]\,\text{d}x\\
&=-H_{e,e'}f(t-t'-\tau)\sum_{\stackrel{\scriptstyle E\subseteq \partial e}{|E|<\theta\,, e'\in E}}\left[\prod_{e''\notin E}\left(1-F_{e''}(t')\right)\prod_{e''\in E\setminus e'}F_{e''}(t')\right]\\
&\quad+H_{e,e'} f(t-t'-\tau)\sum_{\stackrel{\scriptstyle E\subseteq \partial e}{|E|<\theta\,, e'\notin E}}\left[\prod_{e''\in\partial e\setminus E\setminus e'}\left(1-F_{e''}(t')\right)\prod_{e''\in E}F_{e''}(t')\right]  
\end{split}
\end{equation}
Hence the Jacobian of $\mathcal{G}$ in the neighbourhood of the limit $\bm{F}^\star$ can be written 
\begin{equation}
\mathcal{J}_{e,e'}(t,t')=\frac{\partial \mathcal{G}[\bm{F}]_e(t)}{\partial F_{e'}(t')}\Big|_{\bm{F}=\bm{F}^\star}=H_{e,e'}f(t-t'-\tau)\Sigma_{e,e'}(t')\,,
\label{J}
\end{equation}
where
\begin{equation}
\Sigma_{e,e'}(t')=\sum_{\stackrel{\scriptstyle E\subseteq \partial e}{|E|<\theta}}\left(\frac{\prod_{e''\in\partial e\setminus E}\left(1-F^\star_{e''}(t')\right)\prod_{e''\in E}F^\star_{e''}(t')}{\displaystyle\big(1-F^\star_{e'}(t')\big)\mathbb{I}_{e'\notin E}-F^\star_{e'}(t')\mathbb{I}_{e'\in E}}\right)\,.
\end{equation}

In the case of simple contagion processes, we have $\Sigma_{e,e'}(t')\approx 1$ for large $t'$. This leads to the simplified expression $\mathcal{J}=H\otimes \mathcal{F}$, where $\mathcal{F}$ is the integral operator
\begin{equation}
\mathcal{F}[g](t)=\int_0^{t-\tau}f(t-t'-\tau)g(t')\,\text{d}t'\,.
\end{equation}
This operator is made diagonal by a Laplace transform, and hence the eigenfunctions of $\mathcal{J}$ are therefore of exactly the form (\ref{mpF2}) found previously. In particular, the presence of the Hashimoto matrix $H$ in (\ref{J}) implies a contribution proportional to $v_{i\to j}$ as a prefactor to the limiting form of $F_{i\to j}$. Note that although this analysis goes via the simplification found at large $t'$, both the delay $\tau$ and the prefactors $v_{i\to j}$ are the same across the whole time range. 

For complex contagions with thresholds $\theta\geq2$, we have $\Sigma_{e,e'}(t')\approx 0$ for large $t'$, which rules out direct use of the linear analysis outlined above. However, progress can be made with the heuristic $\Sigma_{e,e'}(t')\approx \sigma$, where $\sigma\in\mathbb{R}^+$ depends on $\theta$, but not the edge in question or the delay time distribution. Essentially, the constant $\sigma$ captures the (multiplicative) additional ``difficulty'' for the contagion to spread with higher thresholds. This approximation implies a modification of the timescales obtained in the linear theory, but does not alter the dependence on the network, and hence the relative time to infection is again proportional to the logarithm of the non-backtracking centrality. Figure 4 in the main text supports this claim with numerical evidence. Computing the delay $\tau$ requires the determination of the precise form of $\sigma$. This appears far more challenging and we leave it for future work. 

\begin{figure}
\centering
\includegraphics[width=0.85\linewidth, trim=80 0 80 0]{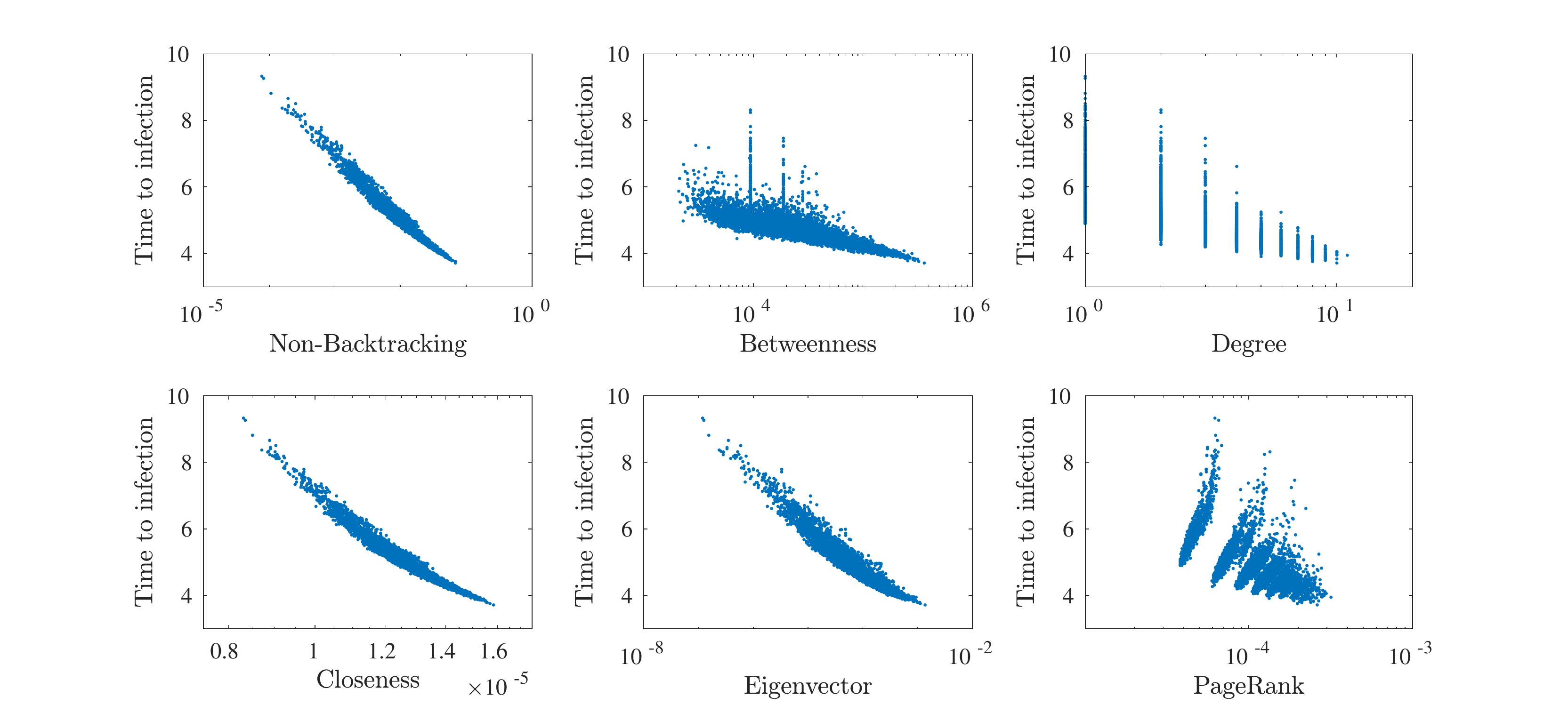}
\caption{Comparison between contagion arival time and (log) centrality of nodes under various metrics. The network used was an Erd\H{o}-R\'enyi random graph on $N=10^4$ nodes with mean degree $3$; contagion simulations were averaged over 1000 samples, with exponentially distributed transmission times.}
\label{fig:centralities}
\end{figure}

\newpage
\supsection{Other centrality measures}
Many different node centrality measures have been proposed in the networks literature, and the weights they ascribe to different nodes are more or less correlated with one another depending on the metrics in question. In Figure~\ref{fig:centralities} we show a comparison between the contagion arival time (as accurately predicted by the non-backtracking centrality) and five other well-known centrality measures. Betweenness centrality, degree centrality and PageRank all show signficant spread of contagion arrival times relavite to centrality score and hence are not useful predictors of, for exmaple,  epidemic risk. 

Closeness centrality is based on the mean distance between nodes in a network, corresponding presicely to the mean contagion arrival time in the limit of Dirac-delta distributed transmission times. Unsurprisingly, this metric shows a strong corelation with epidemic arrival when presented in the logarithmic scale which we have shown to be the correct. We should emphasise three points of advantage our results have over the heuristic use of closeness centrality. First, our theoretical derivations have shown (log) non-backtracking centrality is the correct measure. Second, our analytical results apply to a broad range of infection time distributions. Third, non-backtracking centrality is significantly faster to compute than closeness centrality. 

\begin{figure}
\centering
\includegraphics[width=0.7\linewidth, trim=0 0 0 0]{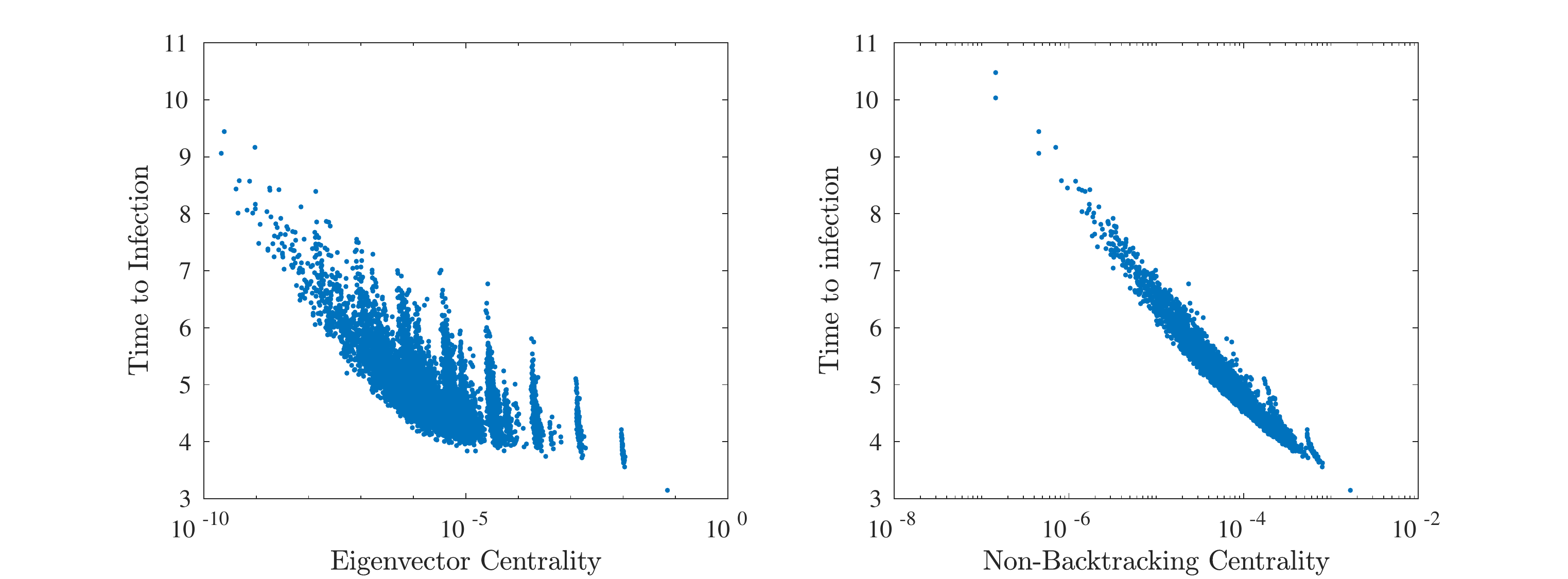}
\caption{Comparison between non-backtracking centrality and eigenvector centrality in predciting infection arival time in the presence of a hub. The network used was an Erd\H{o}-R\'enyi random graph on $N=10^4$ nodes with mean degree $3$, with the addition of a single ``hub'' node connected to 50 randomly selected other nodes. Contagion simulations were averaged over 1000 samples, with exponentially distributed transmission times.}
\label{fig:centralities2}
\end{figure}

In the simple test presented in Figure~\ref{fig:centralities}, the logarithm of eigenvector centrality also appears to make a usable prediction of the infection arrival time. It is well-known, however, that eigenvector centrality can become localized in networks containing high-degree ``hub'' nodes \cite{martin2014localization}. We illustrate this problem in Figure \ref{fig:centralities2}, showing how eigenvector centrality is distorted by the presence a hub node; the striations visible in the left panel correspond to distance from the hub node, which is given undue prominance in eigenvector centrality. Non-backtracking centrality, by contrast, continues to perform well for this example. 


\newpage
\supsection{Temporal percolation}
While the shape of the infection size curve may be described by the non-backtracking centrality distribution of the nodes, one can predict the time at which the bulk of the outbreak occurs by considering the following simple heuristic for unit rate infections. The time needed for an infected cluster of size $m$ to grow by one node scales as $1/E_m$ where $E_m$ is the number of edges leaving the cluster. During the exponential growth phase, addition of a node to the cluster will cause the loss of one external edge (used for transmission) and an average gain of $\lambda$ additional edges from the neighbours of the new node. Therefore $E_{m+1}= E_{m}+\lambda-1$, and we thus find that the time needed to infect a fraction $\beta$ of all nodes grows like $\log(N)/(\lambda-1)$. This implies that rescaling time by $t'=t(\lambda-1/\log(N)$ should collapse the curves for different large networks, as shown in Fig.~5 of the main text. Moreover, the transition between $\beta=0$ and $\beta=1$ becomes sharp at $t'=1$ in the limit of large network size. 

This explosive transition is, in fact, a general property of time-ordered percolation. In the supplement we show how the above reasoning can be made precise for sparse configuation-model networks with finite second moments, corresponding to graphs with non-zero percolation thresholds. Additionally, we study temporal percolation on dense graphs with a fraction $q$ of all possible edges present, showing that the same result holds under the rescaling $t'=tqN/\log(N)$.  

\textbf{A sharp transition in sparse graphs}

Write $R_m$ for the time to reach $m$ nodes. For exponential transmission times
\begin{equation}
\mathbb{E}[R_{m+1}]=\mathbb{E}[R_m]+\mathbb{E}\left[\frac{1}{O_m}\right]\,,
\end{equation}
where $O_m$ is the number of outgoing edges from the cluster of $m$ infected nodes. In a configuration model graph we can write
\begin{equation}
O_{m+1}=O_{m}+K_{m}-1\,,
\end{equation}
where $K_{m}$ is the degree of the node added as the cluster grows from size $m$ to $m+1$, and the $-1$ term corresponds to the edge used in transmission, that becomes internal to the cluster. This leads to the expression
\begin{equation}
O_m=K_0+\sum_{i=1}^{m-1}K_i-(m-1)\,.
\label{sumK}
\end{equation}
For a configuration model graph, $K_0$ is chosen according to the degree distribution and all subsequent $K_i$ are chosen according to the branching distribution. 

Introduce generating functions 
\begin{equation}
g(x)=\mathbb{E}[z^{K_0}]\,,\quad \tilde g(z)=\mathbb{E}[z^{K_1}]\,,\quad \gamma_m(z)=\mathbb{E}[z^{O_m}]\,.
\end{equation}
Now (\ref{sumK}) implies
\begin{equation}
\gamma_m(z)=z^{-m}g(z)\tilde g(z)^{m-1}\,.
\end{equation}
Asymptotically in large $m$ we have 
\begin{equation}
\gamma_m(z)\approx \frac{g(z)}{g'(z)}e^{m(1-z)(\lambda-1)}\,,
\end{equation}
where we used the fact that $\tilde g'(1)=\lambda=1/\rho_c$. It then follows that
\begin{equation}
\mathbb{E}\left[\frac{1}{O_m}\right] = \int_0^1 \gamma_n(z) dz\approx \frac{1}{m(\lambda-1)}\,,
\end{equation}
and hence
\begin{equation}
\mathbb{E}[R_{m}]\approx \frac{\log(m)}{\lambda -1}\,.
\end{equation}

\textbf{A sharp transition in dense graphs}

We show how the sharp transition in infected proportion is not a phenomenon exclusive to this setting but rather a characteristic of time dependent percolation in general by considering a dense network setting also.

Suppose we have for simplicity an infection spreading at exponential rate 1 on a dense network of size $ N $ and mean degree $ qN $, starting from a single node. Taking $ \mathbb{E}(R_m) $ to denote the expected time to reach $ m $ infections, it follows that the time for the number of infected to increase from $ m $ to $ m+1 $ will be given in the early stages by
\begin{align}
\mathbb{E}(T_{n+1}-T_{n})= \frac{1}{qn(N-n)}
\end{align}
and so  $ \mathbb{E}(R_m) $ is given, and bounded, by

\begin{align}
\mathbb{E}(R_m)=&\sum_{\ell=1}^{m-1}\frac{1}{qN\ell-q\ell^2} \nonumber\\
>& \sum_{\ell=1}^{m-1}\frac{1}{q(N-1)}\frac{1}{\ell} = \frac{1}{q(N-1)}\log(m)
\end{align}

For an upper bound first notice that $ \frac{1}{q\ell(N-\ell)}  $ is decreasing with $ \ell $ in the interval $ (0,N/2] $ and so for $ m\leq N/2 $ the sum may be bounded by the integral, so

\begin{align}
\mathbb{E}(R_m)=&\sum_{\ell=1}^{m-1}\frac{1}{q\ell(N-\ell)} < \frac{1}{qN}+\int_1^{m-1} \frac{1}{q\ell(N-\ell)} \, \text{d}\ell\nonumber\\
<&\frac{1}{qN}\left( 1+ \log\left( \frac{(m-1)(N-1)}{N-(m-1)}\right)\right) \nonumber\\
<& \frac{1}{qN}\left( 1+ \log\left( \frac{Nm}{N-m}\right) \right) 
\end{align}

Similarly for $ m>N/2 $ a bound may be made by shifting the end points of the integral to get 
\begin{align}
\mathbb{E}(R_m)<& \frac{4}{qN^2}+\int_{N/2}^{m} \frac{1}{q\ell(N-\ell)} \, \text{d}\ell 
\nonumber\\ <& \frac{1}{qN}\left( 1+ \log\left( \frac{Nm}{N-m}\right) \right) 
\end{align}
(note we are only interested in large $ N $)

These bounds motivate introducing the rescaled variables
\begin{equation}\label{defQ}
Q_\beta=\frac{qN}{\log(N)}T_{\lceil\beta N\rceil}\,,\quad\text{for } \beta\in(0,1)\,.
\end{equation}

Then 
\begin{equation}
1+\frac{\log(\beta)}{\log(N)}<\mathbb{E}Q_\beta <1+\frac{\log(\beta)}{\log(N)}+ \frac{1-\log(1-\beta)}{\log(N)} \,.
\end{equation}

If we rescale time by $t'=qNt/(\log N)$ then, for large $ N $, up to time $t'=1$ almost nobody is infected and after time $t'=1$ the infection reaches almost everyone. 

\begin{figure}
\centering
\includegraphics[width=0.7\linewidth]{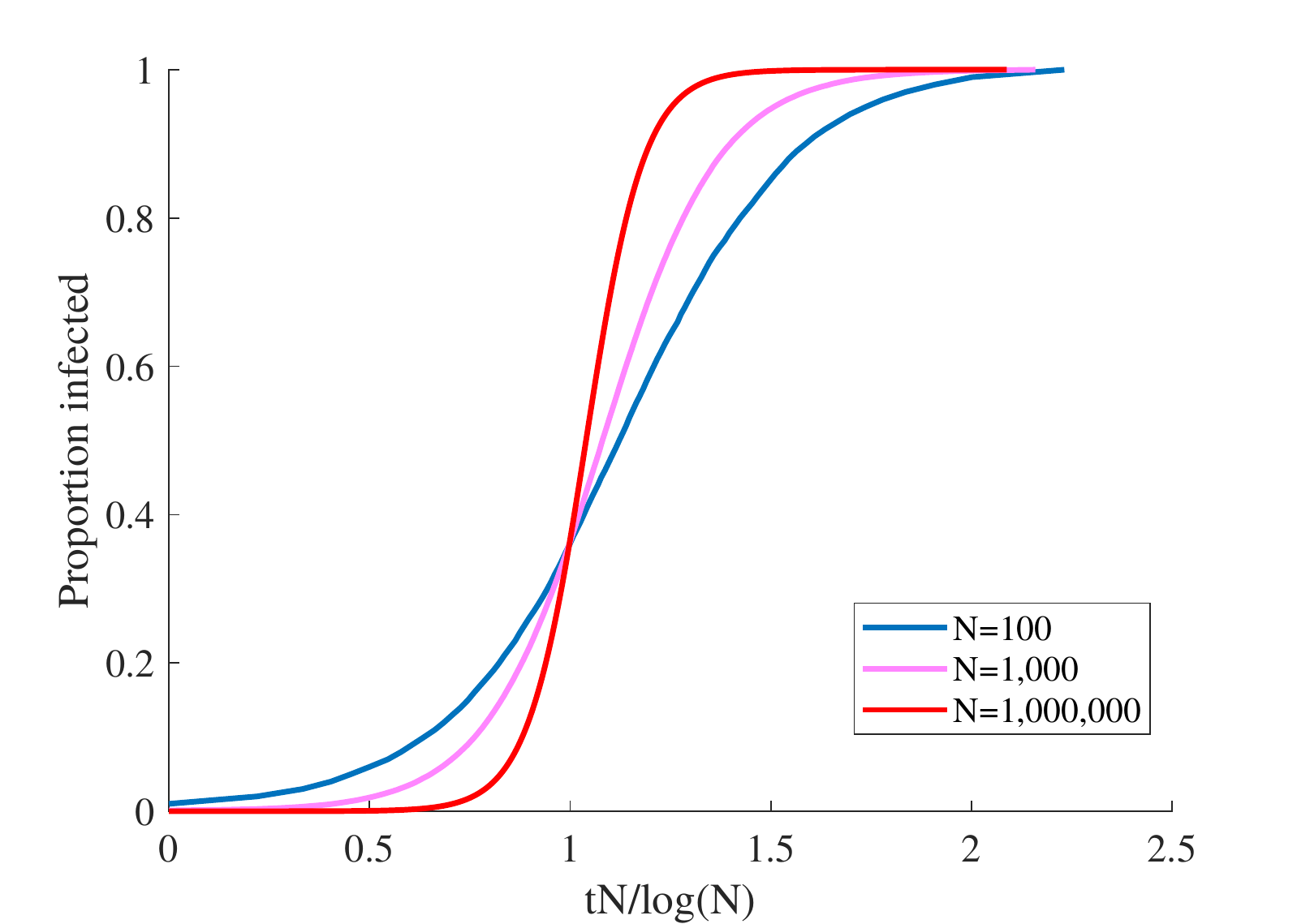}
\caption{Plot showing the proportion of infected vs time on complete graphs of varying size for an infection with exponential infection times as an average over 500 simulations from random seed nodes. Time is scaled by $N/\log(N) $ as per our calculation to show a progressively sharp transition at time 1.}
\label{fig:denscurves}
\end{figure}

\newpage
\supsection{Simulation details}
For a given network $G=(V,E)$, the SI model is simulated as follows:
\begin{enumerate}
 \item Choose a source node $s\in V$ uniformly at random
 \item For each $i\in V$, compute the length of the shortest path from $s$ to $i$, and store this as the distance $d_i$ 
 \item For each ordered pair of neighboring vertices $i$ and $j$ generate a random delay $X_{i,j}$, chosen from the distribution with pdf $f$
 \item For each $i\in V$, compute the minimum weight $w_i$ of a path from $s$ to $i$ in the weighted digraph with weights given by the delays computed in step 3. That is, $$w_i=\min\left\{X_{s,\ell_1}+X_{\ell_1,\ell_2}+\cdots+X_{\ell_m,i}\,:\, (s,\ell_1,\ldots, \ell_m,i)\,\,\text{is a path from $s$ to $i$}\right\}.$$
 \end{enumerate}
Following this procedure, we generate exact samples of the arrival times of the underlying epidemic spreading process. Each sample gives a set of $N$ pairs of the form $(d_i,w_i)$ for $i\in V$, where $d_i$ is the distance from the source and $w_i$ the epidemic arrival time. 

\begin{figure}
\includegraphics[width=0.6\linewidth]{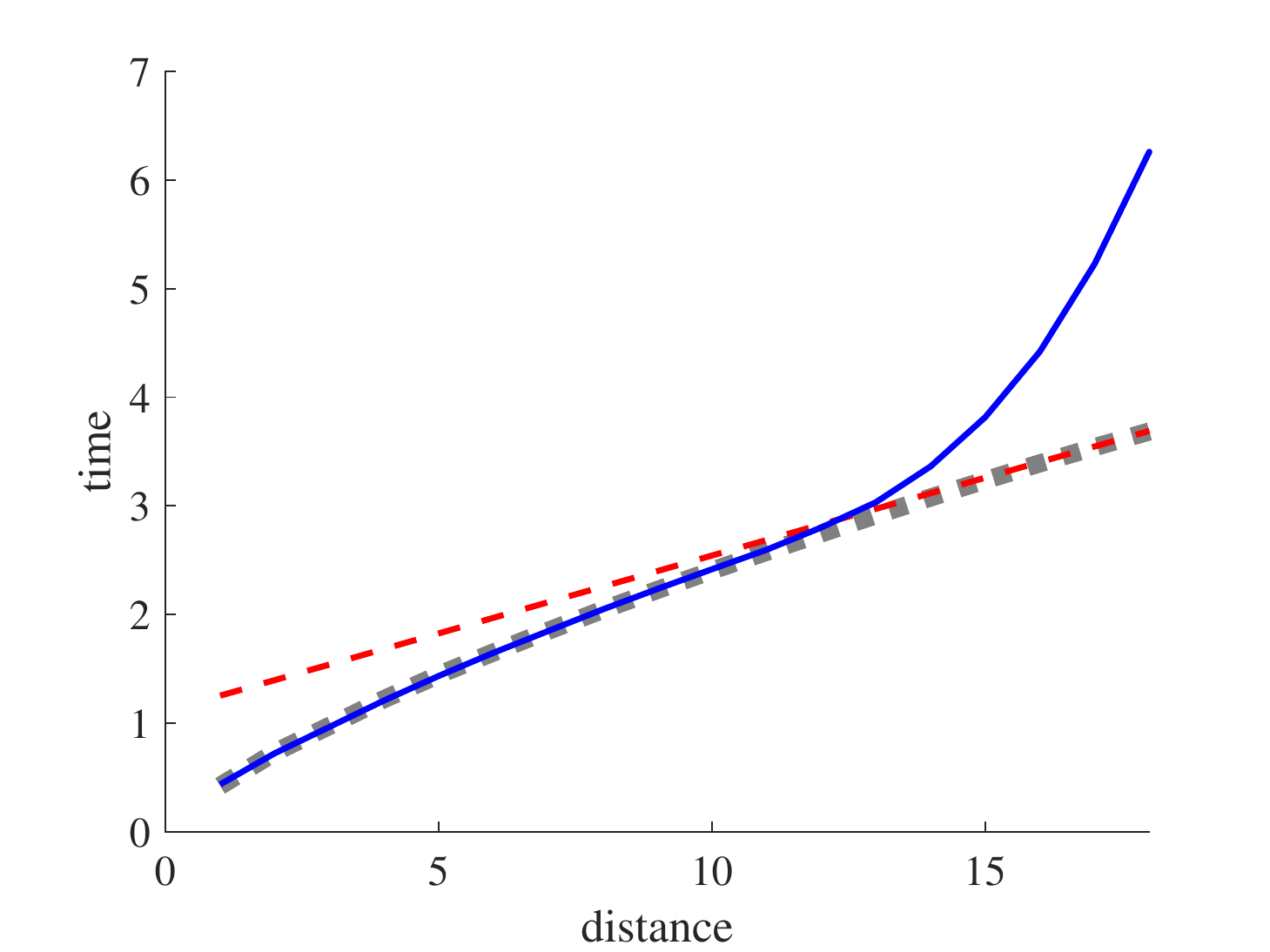}
\caption{Plot of the time $\langle t_n\rangle$ for an infection to reach a given distance $n$ from a random source. For this figure we simulated an infection with exponential unit rate transmission time in an Erd\H{o}s R\'enyi graph of $10^7$ nodes and mean degree $3$, averaged over 100 samples. 
The red dotted line shows the gradient $\tau$ as predicted by our theory for spreading speed.
The grey dashed line shows numerical simulations of a branching process model tuned to match the network in question, effectively removing the large $n$ finite size effects. 
}
\label{fig:tau}
\end{figure}

To compute the expected arrival times for the scatter plots in Fig.~4 of the main text, we simply average the $w_i$ over many samples with different random source nodes. To compute the delay $\tau$ requires additional consideration of the distance from the source. For each $n$, one can compute the time for a simulated contagion to reach that distance from the source by computing $t_n=\min_i\{w_i\,:\, d_i=n\}$. Averaging this quantity over many samples produces $\langle t_n\rangle$, the mean time for a contagion to reach distance $n$ from a random source. Figure \ref{fig:tau} shows a typical example of how this quantity varies with the distance considered. 

In the networks we are interested in, the number of nodes at distance $n$ from the source grows exponentially with $n$ (the small world property). We have developed our theory under the assumption that $n\gg 1$, but still small relative to the diameter of the network. For $n$ too small or too large the speed of spread will be limited by a lack of multiplicity of routes of that length. In this sense the theoretical value of $\tau$ we compute is the smallest possible, corresponding to the regime in which the contagion is spreading rapidly through the bulk of the network. To measure this from simulation data we take $\tau=\min_n\{\langle t_{n+1}\rangle -\langle t_n\rangle \}\,.$

For simulating complex contagions with a threshold $\theta>1$, we find it easier to use an event-based algorithm. Starting from a randomly chosen set of source nodes (larger for higher $\theta$ to give the contagion a chance to take hold), we keep track of individual contagion events to determine for each node when its threshold is reached. 

\supsection{Table of data for main Figure 2}
Table entries are ordered according to their position in the figure, reading right to left.
\vspace{10pt}

\hspace{-1pt}\begin{tabular}{|l|l|l|l|l|l|l|l|}
\hline
Network name & $\lambda$ & $\tau$ & $\tau_{\text{emp}}$ & N & $\langle k\rangle$  & Clustering & Assortativity\\\hline
Deezer Romania \cite{rozemberczki2018gemsec}&16.0&0.0236&0.0225&41773&6.024&0.0912&0.1140\\
Amazon Co-purchasing \cite{leskovec2007dynamics}&17.8&0.0211&0.0188&262111&6.866&0.4198&-0.0025\\
Amazon Products \cite{yang2015defining}&20.7&0.0181&0.0162&334863&5.530&0.3967&-0.0588\\
Deezer Hungary\cite{rozemberczki2018gemsec}&22.9&0.0163&0.0187&47538&9.377&0.1162&0.2072\\
Facebook Companies \cite{rozemberczki2018gemsec}&30.6&0.0122&0.0145&14113&7.387&0.2392&0.0126\\
Arxiv Cond. Mat.\cite{leskovec2007graph}&35.8&0.0104&0.0145&21363&8.546&0.6417&0.1253\\
Facebook Athletes \cite{rozemberczki2018gemsec}&44.3&0.0084&0.0124&13866&12.521&0.2762&-0.0270\\
Deezer Croatia \cite{rozemberczki2018gemsec}&46.2&0.008&0.0104&54573&18.258&0.1365&0.1971\\
Brightkite Social \cite{cho2011friendship}&99.9&0.0037&0.0086&56739&7.5&0.1734&0.0096\\
Enron Email \cite{leskovec2009community}&115.5&0.0032&0.0055&33696&10.732&0.5092&-0.1165\\
Epinions \cite{richardson2003trust}&181.6& 0.0021 &0.0044&75877 &10.695&0.1378&-0.0406 \\
School contact \cite{salathe2010high}&335.9 &0.0032 &0.00099&788&300.2&0.499&0.0539\\\hline
\end{tabular}

\end{document}